%% file: Q_Diffusive_Monitoring.tex
\documentclass[aps,pra,twocolumn,showpacs,superscriptaddress]{revtex4-1}  
\usepackage{graphicx}  
\usepackage[caption=false]{subfig}
\usepackage{float}
\usepackage[toc,page]{appendix}
\usepackage[table]{xcolor}
\usepackage{dcolumn}   
\usepackage{bm}        
\usepackage{amssymb}   
\usepackage{eufrak}   
\usepackage{color}

\usepackage{amsmath, amsthm, amssymb}
\usepackage{amsfonts}
\usepackage{bbold}       
\usepackage{accents}    
\newcommand{\ut}{\underaccent{\tilde}}
\usepackage{soul}
\newcommand{\beq}{\begin{equation}}
\newcommand{\eeq}{\end{equation}}
\newcommand{\bqa}{\begin{eqnarray}}
\newcommand{\eqa}{\end{eqnarray}}

\newcommand{\erf}[1]{Eq.~(\ref{#1})}

\newcommand{\erfs}[2]{Eqs.~(\ref{#1})--(\ref{#2})}
\newcommand{\erfa}[2]{Eqs.~(\ref{#1}) and (\ref{#2})}
\newcommand{\arf}[1]{{ App.}~\ref{#1}} 
\newcommand{\srf}[1]{Sec.~\ref{#1}} 
\newcommand{\crf}[1]{Ref.~\cite{#1}} 
\newcommand{\trf}[1]{Table ~\ref{#1}} 

\newcommand{\frf}[1]{Fig.~\ref{#1}}

\newcommand{\ie}{{\it i.e.}}
\newcommand{\dg}{^\dagger}

\definecolor{BLACK}{gray}{0}
\definecolor{RED}{rgb}{1,0,0}
\definecolor{GREEN}{rgb}{0.2,.6,0.2}
\definecolor{BLUE}{rgb}{0,0,1}

\newcommand{\blk}{\color{BLACK}}

\newcommand{\sch}{Schr\"odinger}

\newcommand{\sq}[1]{\left[ {#1} \right]}

\newcommand{\an}[1]{\left\langle{#1}\right\rangle}

\newcommand{\abs}[1]{\left| {#1} \right|}

\newcommand{\s}[1]{\hat \sigma_{#1}}

\newcommand{\tp}{^{\top}} 

\begin{document}

\widetext


\title{Detector dependency of diffusive quantum monitorings}
\input author_list.tex       
\date{\today}

\begin{abstract} 
 Continuous \blk measurements play a pivotal role in the study of dynamical open quantum systems. `Dyne' detections are among the most
widespread and  efficient \blk measurement schemes,  and give rise to quantum diffusion of the conditioned state.
\blk In this work we study  under what conditions the \blk detector dependency of the conditional state of a quantum system subject to 
diffusive  monitoring can be demonstrated experimentally, in the sense 
of ruling our any detector-independent pure-state dynamical model for the system.  We consider \blk an arbitrary number $L$ of environments to which the system is coupled,  and an arbitrary number $K$ of different types of dyne detections.  \blk  We prove that 
 non-trivial necessary conditions for such a demonstration \blk can be  determined efficiently by
semi-definite programming.  To determine sufficient conditions, \blk different physical  environmental couplings and Hamiltonians for a qubit, and different  sets of diffusive monitorings are \blk scrutinized.  We compare the threshold efficiencies that are sufficient in the various cases, as well as cases previously considered in the literature, to suggest the most feasible experimental options.

\end{abstract}

\maketitle

\section{Introduction}
One of the most challenging tasks in the analysis of an open quantum system is to gain as much information as 
possible about the system state. The conventional approach of tracing over the environment, with which the system
is coupled to, would not yield to the maximal knowledge. However, continuous measurements (also referred to as
{\it monitorings}) of the environment make it possible to gain further information about the system by exploiting 
quantum correlations between the two \cite{WisMil10}. The former approach results in a master equation (ME) for the state matrix $\rho$
of the system whereas the latter gives a stochastic master equation (SME) describing evolution of a conditional
state matrix $\varrho$ given a readout.  The latter is \blk known as an {\it unraveling} 
of the ME \cite{Car93},  because the ME is recovered by averaging over all possible measurement \blk records.  Under ideal conditions, \blk for every single measurement record there is a  path of 
a  ray in the system's Hilbert space, \blk called a {\it quantum trajectory}. 

 A \blk SME  may have more than \blk one physical realization (that is, the set of
different ways of monitoring the environment is  a surjective mapping onto \blk the set of all possible SMEs).
In the context of quantum optics, different monitorings of the environment can be achieved by adding a local oscillator 
(LO) to the system output field prior to detection. Depending on the strength,
frequency, and phase of the LO in comparison with those of the emitted radiation the dynamics of the system
are described by distinct SMEs.   These \blk fall into two categories: jump unravelings and diffusive 
unravelings \cite{WisMil10}. For quantum jumps the LO either is a weak field or is set at zero (direct detection), while
for quantum diffusion it is  a strong field, which can be \blk resonant with the output field (homodyne detection) 
or detuned from  it \blk (heterodyne detection) \cite{WisMil10,BacRal04}.


In this paper we scrutinize diffusive unravelings of MEs  in the context of experimentally 
demonstrating the \blk detector dependency of  quantum dynamics \blk \cite{WisGam12}. Historically, before
the emergence of quantum trajectory theory \cite{Car93,DalCasMol92,GarParZol92,Bar93,WisMil93}, it was 
 widely believed \blk that a system's quantum dynamics was objective, and thus independent of any measurement. 
This notion was proposed in early works of Einstein on emission and absorption of light quanta by an atom \cite{Ein17} 
such that the system (atom) state undergoes evolution among stationary states of Bohr \cite{Boh13}.
But since the early 90's, modern theory of quantum dynamics has rendered this idea obsolete, 
 implying \blk that dynamical quantum events are subject to a remote detector. 

 It remains a challenge, however, to \blk verify experimentally the testimony of this theory.  
 In 2012, a  proposal was made for an experiment \blk to disprove all objective
pure-state dynamical models (OPDM) for  a particular quantum system \blk ~\cite{WisGam12}. \blk 
 If an experimentalist had  perfect detection (and collection) \blk efficiency, $\eta=1$, 
 the conditioned pure state would be pure, and it would simply be a matter of verifying that 
two different detection schemes lead to different types of pure states. That is, the problem would be \blk a continuous-in-time version of the 
Einstein-Podolsky-Rosen (EPR) phenomenon \cite{EPR35}.
However, in the real experimental
setups $\eta < 1$, and in order to rule out all OPDMs one would require to invoke the more general form of the
EPR argument  known as steering~\cite{WisJonDoh07} (the term introduced by \sch~for this phenomenon~\cite{Sch35}), 
or EPR-steering~\cite{CavRei09}. \blk In this formalism one experimenter
(Alice) proves that the objective quantum state of the other experimenter (Bob) cannot be explained by any local hidden state (LHS) models. 
This is fulfilled by expressing the correlation of measurements of the two parties in terms of an inequality, the violation of which can be interpreted
as ruling out all  LHS models for Bob's system. Generalizing this to continuous monitoring~\cite{WisGam12}, 
inequalities can be derived of which the violation would rule out all OPDMs, and consequently confirm the 
\blk detector-dependency of stochastic quantum dynamical evolutions. 

 The physical system studied in \crf{WisGam12} was a resonantly driven two-level atom
with two measurement arrangements to capture the fluorescence. It turns out that using two homodyne detection,
and assuming both have equal efficiencies,  the \blk critical efficiency  to
disprove OPDMs  is $\eta_c \approx 73\%$, which is quite high. \blk 
Further, it was proved~\cite{WisGam12} that $\eta > 50\%$ is a prerequisite 
for being able to demonstrate EPR-steerability using any dyne detection schemes.
 Very recently, two of us have generalized this necessary condition~\cite{DarWis14}, 
showing that \blk for any Markovian open quantum system with an arbitrary number $L$ of decoherence 
channels to the environment and with an arbitrary number $K$ of diffusive unravelings,  unless at least one member of the set 
$\{\eta_{\ell}^k\}$ is greater than $50\%$,  the diffusive \blk unravelings cannot be used to demonstrate detector dependency of the conditional state
matrix $\varrho$.

The present work  furthers the investigation of how well the \blk detector dependency of dynamical quantum events  
can be proven experimentally \blk using diffusive
 unravelings,  in terms of the efficiency required. 
 Our first aim is to derive even more general necessary conditions, and our second aim is to derive 
 sufficient conditions in a variety of scenarios with a simple system (a qubit). \blk 
 
 We organize the paper as follows. First, \srf{sec:II} reviews briefly formalism of general
 diffusive unravelings. In \srf{sec:III} it is then shown that the most general proof of a no-go for inefficient diffusion 
 can be  cast as a special case of \blk a semidefinite programming (SDP)   
 known as a {\em feasibility problem} (FP). We present a simple  graphical representation 
 of this general necessary condition \blk for 
 the case where the system interacts with just one environment. Following this, In \srf{sec:IV} different 
  MEs and \blk measurement 
 strategies are considered.  First \blk we analyze a system with three irreversible channels, 
  and three different diffusive unravelings to try to prove detector-dependence. \blk Next, systems with  fewer 
 \blk decoherence channels and just two  different \blk diffusive unravelings are studied. Finally, in \srf{sec:V} we 
 conclude by summarizing the results of this study
along with  the \blk outcomes  of previous publications \blk in \trf{tbl:S}.

\section{General diffusive unraveling with imperfect detection} \label{sec:II}
The state $\rho$ of an open quantum system in the Markov approximation,  defined \blk by tracing over the 
environment,  has its evolution \blk governed by a ME
\beq \label{ME}
\dot{\rho} = -i\big[\hat{H}, \rho\big] + {\cal D} [\hat{\bf c}] \rho \equiv {\cal L} \rho.
\eeq
Here $\hat{H}$ is the system Hamiltonian  (in units where $\hbar = 1$),  \blk
$\hat{\bf c} = (\hat{c}_1,\ldots,\hat{c}_L)^{\top}$ is a vector of arbitrary
operators (also know as Lindblad operators \cite{Lin76}), 
and ${\cal D} [\hat{\bf c}] = \sum_{l=1}^{L} {\cal D} [{\hat{c}}_l]$ where the decoherence 
superoperator ${\cal D} [\hat{o}] \rho = \hat{o} \rho {\hat{o}}\dg -1/2\{{\hat{o}}\dg \hat{o}, \rho \}$, 
with curly brackets representing the anticommutator.  

The most general diffusive unraveling of the ME, \erf{ME}, by allowing for inefficient detection reads \cite{WisMil10}
\beq \label{SME}
{d\varrho} = {\cal L} \varrho \,dt+ {\cal H} [d{\bf V}\dg \hat{\bf c}] \varrho, 
\eeq
where ${\cal H} [\hat{o}] \rho = \hat{o} \rho + \rho {\hat{o}}\dg - {\text Tr} [\hat{o} \rho + \rho {\hat{o}}\dg] \rho$
is a nonlinear superoperator, and
$d{\bf V} = (dV_1,\ldots,dV_L)^{\top}$ is a vector of infinitesimal c-number Wiener increments. 
These are Gaussian random variables satisfying ${\text E}[d{\bf V}] = 0$, where ${\text E}[\diamond]$
denotes the ensemble average of the random variable $\diamond$ with respect to its probability distribution function.
Physically, they should actually be regarded as noise in the output complex photocurrent 
\beq \label{Ccurrent}
{\bf J}\, dt= \an{{\Theta} \, \hat{\bf c} + \Upsilon \, \hat{\bf c}^{\ddagger} } dt + d{\bf V},
\eeq
with the following correlation relations
 \beq \label{Wiener_cor}
d{\bf V} d{\bf V}\dg = {\Theta} dt,  \;\;\;\;   d{\bf V} d{\bf V}^\top = {\Upsilon}\, dt,
\eeq
where $ {\bf o}^\ddagger = ({{\bf o}^{\top}})\dg$, and $\an{\hat{o}} = {\rm Tr}[{\hat{o}} \varrho]$ 
is the quantum mechanical 
expectation value of an operator $\hat{o}$  with respect to the  conditioned quantum state $\varrho$. \blk
Here $ \Theta = {\rm diag}(\eta_1,\ldots,\eta_L)$ is a real diagonal matrix allowing for imperfect detection in the
formalism so that channel $l$ is monitored with  efficiency \blk $0 \leq \eta_l \leq 1$. The rest of parameters that characterize
diffusive unravelings are being encoded in a complex symmetric matrix ${\Upsilon} = {\Upsilon}^\top$. It
is convenient to encapsulate all of the correlation properties in the so-called {\em unraveling matrix} \cite{WisDio01}
\beq  \label{formU}
U({\Theta},{\Upsilon}) \equiv  \frac{1}{2}\left( 
\begin{array}{cc}
{\Theta}+\text{Re}\left[ {\Upsilon} \right]  & \text{Im}\left[ {\Upsilon} \right]  \\ 
\text{Im}\left[ {\Upsilon} \right]  & {\Theta}-\text{Re}\left[ {\Upsilon} \right] 
\end{array}
\right),
\eeq
with the only constraint that it must be positive semi-definite (PSD), \ie\ that
 $\exists Z\in \mathbb{C}^{2L\times 2L}$ such that $U({\Theta},{\Upsilon}) = Z^\top Z$.
 Of course, one could work with
 other matrix representations of diffusive measurements other than $U$ \cite{ChiWis11}, but for the sake of consistency
 with \crf{DarWis14} we rather keep this notation, even though we shall also give a technical reason why it is not an 
 appropriate choice (see \srf{sec:IIB}).
 Thus an ideal monitoring of the environment giving rise to a pure-state quantum diffusion
 is represented  by $U(I, \Upsilon)$, \blk where $I \in {\mathbb R}^L$ is an identity matrix, and correspondingly \erf{SME} 
 can be \blk replaced with a stochastic Schr\"dinger equation.

 
An interesting class of unravelings is obtained when the  Lindblad \blk operators are Hermitian $\hat{\bf c} =  \hat{\bf c}^\ddagger$ and the complex Wiener increments
 satisfy
 \beq   \label{W:ML}
dV_l  =  e^{i \phi} \sqrt{\eta} \, dW_l, \quad l= 1, \cdots, L,
\eeq
where
\beq \label{Wien_inc1}
d W_j\,d W_k =  \delta_{j k}\, dt.
\eeq
Therefore, depending on the monitoring scheme the  $l$th \blk measurement outcome would be either a current 
 containing maximal information about the observable $\hat{c}_l$, \blk 
\beq \label{J:current}
J_l^{\rm current} dt =  2 \eta \an{\hat{c}_l} dt + \sqrt{\eta}\, dW_l  \;\;\; (\phi = 0),
\eeq
or just a pure-noise, 
\beq \label{J:noise}
J_l^{\rm noise} dt =  i \sqrt{\eta} \,dW_l  \;\;\; (\phi = \pi/2).
\eeq
 In the first case, the associated conditional evolution tends to localize the system to a $\hat{c}_l$ eigenstate: 
\beq
d \varrho = {\cal L} \varrho \,dt + \sqrt{\eta} \sum_{l=1}^L \big[(\hat{c}_l - \an{\hat{c}_l} ) \varrho + {\rm H.c.} \big] dW_l . 
\eeq
In the latter case, the stochastic evolution corresponds 
to a noisy Hamiltonian, since no information about the system is being obtained:
\beq
d \varrho = {\cal L} \varrho \,dt + i \sqrt{\eta} \sum_{l=1}^L \big[\varrho , \hat{c}_l \big] dW_l . 
\eeq
 We employ these kind of unravelings in \srf{sec:IV} for some specific \blk scenarios.
 
 Now having this formalism in our arsenal we  can investigate proving \blk detector dependency of  diffusive \blk stochastic evolution. 
 
\section{Necessary conditions for demonstrating detector dependence} \label{sec:III}

\subsection{Coarse graining} 

 In addition to \blk the mathematical tools that have been described thus far we require another
essential concept, that is,
{\em coarse graining} of diffusive unravelings. To this end, consider two unravelings $U \equiv U({\Theta},{\Upsilon})$ and 
$U_0 \equiv U({\Theta}_0,{\Upsilon}_0)$ with associated vectors
of Wiener processes $d{\bf V}$ and $d{\bf V}_0$, respectively. If it is possible to express the latter as the
algebraic sum of the former and another complex vector Wiener increment $d\check{{\bf V}}$
 obeying the above restrictions, \ie\ $d{\bf V}_0 = d\check{{\bf V}} + d{\bf V}$, then the first unraveling $U$ can be realized experimentally by
implementing the second one, $U_0$, and keeping just  the \blk relevant information in record, 
 discarding the information in  $d{\bf V}$. \blk 
This will be so if the tacit unraveling
matrix for $d\check{{\bf V}}$, $\check{U} \equiv U_0 - U$, is also PSD. Under these conditions 
we call $U$ a coarse graining of $U_0$,  and $U_0$ a fine graining of $U$ (and of $\tilde{U}$). \blk 

\subsection{Arbitrary number $L$ of environments} \label{sec:IIB}
 Consider \blk an experimenter  who \blk is able to implement a set $\mathfrak U$ of  some \blk 
number $K>1$ of distinct unravelings 
\beq \label{U_set}
{\mathfrak U} = \{ U_k \equiv U({\Theta}_k,{\Upsilon}_k): U_k \geq 0, k=(1,\ldots, K) \}.
\eeq
As pointed out in the introductory section,  it is necessary to perform EPR-steering in order to prove that \blk  conditional
stochastic evolution of the system is  determined by \blk a distant detection apparatus. 
 This means that \blk if there exists a single unraveling $U_0$ so 
that $\forall \, k, \check{U}_k \equiv U_0 - U_k \in {\mathbb R}^{2L\times 2L}$ is PSD,  then every member of \blk
the set ${\mathfrak U}$ 
 is a coarse graining \blk of $U_0$.   If such a $U_0$ exists, then $\Theta_0$ can always be chosen 
to equal $I$ (as this just makes every $\tilde{U}_k$ more positive), so that $U_0$ is a purity-preserving unraveling. 
In other words, there exists a pure-state dynamical model which could underly all of the different 
stochastic evolutions induced by the $K$ different measurement schemes, namely that corresponding to $U_0$. 
Consequently, that set $\mathfrak U$ cannot possibly be used to rule out all OPDMs. 

Thus we have a necessary condition for demonstrating detector dependence: the non-existence of a 
common fine-graining $U_0$ for every member of $\mathfrak U$. But is there a way to 
establish the existence, or non-existence, of such a $U_0$ given $\mathfrak U$? \blk


In \crf{DarWis14}  a partial answer was given, in the form of \blk a no-go theorem for inefficient diffusive unravelings: if for all decoherence channels $l$ and for all monitoring schemes $k$ the detection efficiencies satisfy
$\eta_l^k \leq 0.5$, then there exists an unraveling $U_0$ from which the set ${\mathfrak U}$ can be obtained by coarse graining. \blk 
This was proven by considering the choice $ U_0 = U(I,0)$.
This  unraveling corresponds to ``quantum state diffusion'' (QSD) as introduced in Refs.~\cite{GisPer92a,GisPer92b} 
(as an objective pure state dynamical model) and could be realized by unit-efficiency heterodyne detection~\cite{WisMil93}. \blk


In this paper  we prove a much more general no-go theorem. Rather than considering a particular $U_0$, and seeing what that implies, we consider any given set ${\mathfrak U}$ and define a procedure to determine whether any $U_0$ exists. 
The procedure is, in general, numerical rather than analytical, but it can be performed efficiently because of the power 
of semi-definite programming~\cite{VanBoy96}. \blk To reiterate the problem definition,  given a set of diffusive unravelings 
${\mathfrak U}=\{{U}(\Theta_k,\Upsilon_k)\}_{k=1}^K$, 
we want to determine whether there exists a  $\Upsilon_0$ such that $U(I,\Upsilon_0)\geq 0$ and 
$\forall k, U(I-\Theta_k,\Upsilon_0-\Upsilon_k) \geq 0$.  \blk

\subsubsection*{Solution via Semi-Definite Programming}

With this formulation in hand, the problem can be formalised as a particular class of semidefinite programming (SDP) 
which is referred to as a {\em feasibility problem} \cite{ParLal03}, the description of which comes below. The standard
 definition of a (dual) semidefinite program is~ \cite{VanBoy96} 
\bqa \label{sdp}
{\rm minimize}       & \;\;\;\;\;\;   &  {\bf b}^\top {\bf x}    \\
{\rm {subject \; to}} & \;\;\;\;\;\;  & F({\bf x}) \geq 0,  \label{LMI1}
\eqa
where
\beq  \label{sdpF}
F({\bf x}) = F^0 + \sum_{j=1}^{n} {\bf x}_j F^j.
\eeq
Here ${\bf b} \in {\mathbb R}^{n}$, and $F_0,\cdots,F_n \in {\mathbb R}^{m\times m}$ are the known data, 
and the minimization is over ${\bf x} \in {\mathbb R}^{n}$.  In terms of these variables, 
the constraint (\ref{LMI1}) is \blk a linear matrix inequality (LMI). In the special case in which ${\bf b}=0$, 
the optimization problem reduces to the search for some vector ${\bf x}$  satisfying \blk 
the LMI, which should obviously be an easier computational job in practice. This is an instance of a {\em feasibility problem} 
and  is exactly \blk what we need. 

 To apply the above 
formalism, we need a one-to-one linear mapping between the symmetric complex \blk $L\times L$ matrix $\Upsilon_0$ and a 
real vector ${\bf x} \in {\mathbb R}^{L (L+1)}$ \cite{ft1} 
\blk
\beq
\Upsilon_0= \heartsuit( {\bf x}) ,
\eeq
Defining $\spadesuit({\bf x})=U(I,\heartsuit( {\bf x}) )$, which is still a linear function of ${\bf x}$, 
the required constraints can be written as the LMI (\ref{LMI1}) by choosing $F({\bf x})$ to be the following  block-diagonal \blk matrix 
$\in {\mathbb S^{2L(K+1)}}$:
\beq  \label{formF}
F({\bf x}) = \left( 
\begin{array}{c|c|c|c}
\cellcolor{gray!15}{\spadesuit({\bf x})}  & 0                           & \cdots & 0  \\ [4pt] \hline 
 & \cellcolor{gray!15}			&				 & \\ 
0                                                & \cellcolor{gray!15}\spadesuit({\bf x}) - U_1 & \cdots & 0  \\ [4pt] \hline
 & &\cellcolor{gray!15} & \\ 
\vdots          & \vdots                    & \cellcolor{gray!15}\ddots & 0  \\[4pt] \hline
 & & &\cellcolor{gray!15} \\ 
0                 & 0                            & \cdots &  \cellcolor{gray!15} \spadesuit({\bf x}) - U_k     
\end{array}
\right).
\eeq
 This is because \blk positive semidefiniteness of \erf{formF} ensures that all 
submatrices must be PSD. Also the linearity of $\spadesuit({\bf x})$ and consequently that 
of $F({\bf x})$ implies that the latter can be written in the form of \erf{sdpF}, with  
 $F^0\cdots, F^n \in {\mathbb S}^{2L(K+1)}$, where $n=L(L+1)$. \blk Therefore the task is to find the following {\em feasible set}
\beq \label{fzb_set}
{\mathfrak F} = \{{\bf x} \in {\mathbb R^{L (L+1)}}: F({\bf x}) \ge 0 \}.
\eeq

Each member ${\bf x}$ of this set is called a {\em feasible point}, and  defines a fine graining 
$\spadesuit({\bf x})$ of every member of the sets ${\mathfrak U}$ as coarse grainings of it. \blk
This FP has $ 2L(K+1)$ constraints with $L (L+1)$ unknown real parameters. 
This suggests that other representations of diffusive unravelings which deal with a larger number of 
unknowns, for example the one introduced in \crf{ChiWis11}, are not computationally efficient for the particular kind of problems
we are considering here.
It \blk is known that a SDP can be solved in (approximately) polynomial time 
with respect to $n=2L(K+1)$, and with errors as small as one wishes \cite{GarMat12}.   Thus, \blk 
finding the unknowns
 would not be costly from computational viewpoints \cite{KulDhi09}, in particular  because \blk no optimization is performed.
If it turns out that the set $\mathfrak F$ is null then all OPDMs will be ruled out.

 This is the no go theorem. It is easiest to see how it works by considering the simple case $L=1$; 
that is, when \blk the system is allowed to couple with 
just one environment.

\begin{figure}
\captionsetup[subfigure]{labelformat=empty}
\subfloat[]{\includegraphics[scale=0.46]{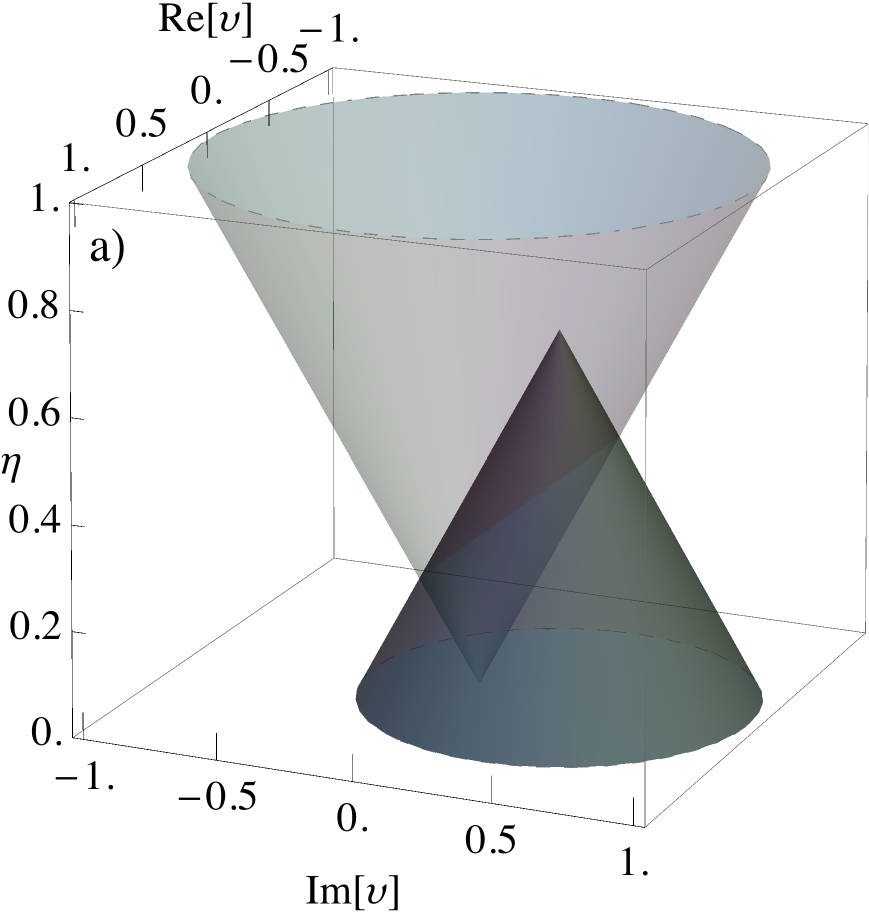} \label{fig:U0G} } \,
\subfloat[]{\includegraphics[scale=0.46]{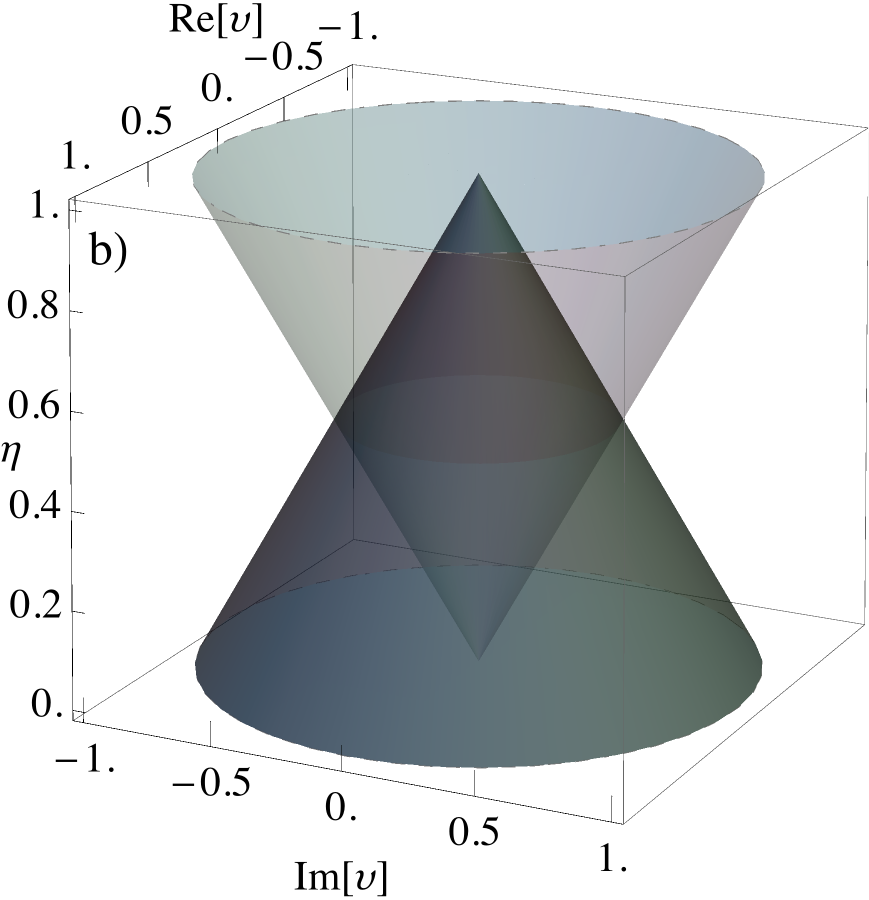} \label{fig:U0}} 
\caption{\label{fig:jump} Feasibility set and space of unravelings for an open quantum system with one environmental decoherence 
channel $L=1$. The light gray cones represent all physical unravelings that satisfy \erf{consU0}, and the dark gray cones show
unravelings that meet the constraints given in \erf{consU0Ukb} for (a) a generic unraveling $U_0=U(\eta_0, { \upsilon}_0)$, and 
(b)   the QSD unraveling \blk
$U_0=U(1,0)$. The intersection of these two conical spaces encloses a volume so that all points inside of it can be a coarse graining of the 
$U_0$ located at the vertex. } 
\end{figure}
\subsection{Single environment $L=1$}
There are numerous physical problems that are modelled using the ME given in \erf{ME} with only one decoherence channel,  such 
as \blk a two-level system with dipole-coupling to a coherent \blk electromagnetic field \cite{MabDoh02}.
In this case the vector of arbitrary operators $\hat{\bf c}$ in \erf{SME} is replaced with a single Lindblad operator $\hat c$ and hence the diffusive unraveling
 is described by
\beq  \label{formU_L1}
U({\eta},{\Upsilon}) \equiv  \frac{1}{2}\left( 
\begin{array}{cc}
{\eta}+\text{Re}\left[ {{ \upsilon}} \right]  & \text{Im}\left[ {{ \upsilon}} \right]  \\ 
\text{Im}\left[ {{ \upsilon}} \right]  & {\eta}-\text{Re}\left[ {{ \upsilon}} \right] 
\end{array}
\right),
\eeq
where $\eta \in {\mathbb R}$ and ${ \upsilon} \in {\mathbb C}$ are scalars corresponding to 
matrices $\Theta$ and $\Upsilon$ in \erf{formU}, respectively. It is straightforward to 
check out the positive semidefiniteness constraint for \erf{formU_L1} by inspecting its spectrum. This constraint
translate to 
\beq \label{consU0}
| { \upsilon} | \leq \eta.
\eeq
Also, the same procedure applies 
to the $K+1$ constraints of the feasibility set, \erf{fzb_set}, so that they all can be arranged in the following form
 \begin{subequations} \label{consU}
\bqa  
| { \upsilon}_0 | & \leq & \eta_0,  \label{consU0Uka} \\
\forall k \qquad | { \upsilon}_0 - { \upsilon}_k | & \leq & \eta_0 - \eta_k  . \label{consU0Ukb}
\eqa
\end{subequations}
Since each diffusive unraveling can be uniquely determined by $(L^2+2L)|_{L=1} = 3$ variables, it is then possible to visualize a feasible set
in three dimensional space such that the coordinates are labeled by ${\rm Re}[{ \upsilon}], {\rm Im}[{ \upsilon}],$ and $\eta$. 
The set of all physically meaningful unravelings that satisfy \erf{consU0} is shown in \frf{fig:U0G} as the light gray cone. 
Therefore the set ${\mathfrak U}$ given by \erf{U_set} that is supposed to be implemented by an experimenter is 
represented by a set of points $\{(\eta_k, { \upsilon}_k)\}_{k=1}^{K}$ in this cone. Moreover, it is evident that unravelings $U_0$
that comply with the condition of \erf{consU0Uka} and eventually make up the feasibility set have to lie somewhere in this conical 
volume.  Now by inspection, for a given $U_0$, \erf{consU0Ukb} states that the points in ${\mathfrak U}$ lie in an 
{\em inverted} cone whose apex is $(\eta_0, { \upsilon}_0)$. \blk Thus, referring to \frf{fig:U0G}, the question now is: 
does there exist a point $(\eta_0, { \upsilon}_0)$ in the light gray cone  which is the apex of an inverted (dark gray) cone that 
includes all the points in ${\mathfrak U}$? \blk 

 If we consider 
$U_0 = U(1,0)$ (heterodyne detection, or QSD), \blk the two cones intersect each other right at $ |\upsilon|=\eta=0.5$; 
see \frf{fig:U0}.
It can obviously be seen that  a set ${\mathfrak U}$ containing only \blk unravelings with efficiencies $\eta \leq 0.5$  is \blk 
not capable of demonstrating EPR-steering (as they are confined 
in the shared volume  of the two cones). \blk In other words, to be able to rule out all OPDMs, at least one efficiency has to be greater than $0.5$. However, as mentioned in \crf{DarWis14} this is just a necessary condition and is not sufficient to prove the detector dependency of stochastic conditional states $\varrho$.
This insufficiency can be seen immediately by looking at the upper half part of \frf{fig:U0} where  there are other \blk unravelings in the overlapped volume  that are also \blk a coarse
graining of  $U(1,0)$, those with $|\upsilon|<1-\eta<0.5$.  

\section{demonstrating detector dependence} \label{sec:IV}

 Having determined necessary conditions for being able to demonstrate detector-dependence using 
diffusive unravelings, we now turn to sufficient conditions. This will involve considering particular 
open qubit systems, coupled to one or more environments, with two or more different unravelings, 
as well as considering various EPR-steering inequalities that might be violated. \blk



\subsection{$L=3$, $K=3$} \label{sec:IV1}

Consider  first \blk an open qubit system with $L=3$ irreversible environmental channels, 
 with the \blk conditional evolution of system's state characterized by the following SME 
\beq \label{SME:M3}
{d\varrho} = \sum_{l=1}^{3} \Big( {\cal D}[\sqrt{\gamma} {\hat{\sigma}}_l] \varrho \, dt+ {\cal H} [\sqrt{\gamma} {\hat{\sigma}}_l dV_l] \varrho \Big).
\eeq
 Here \blk ${\hat{\sigma}}_l$ are the usual Pauli operators. In comparison to the notation of \erf{SME},
$\hat{H}=0$ in a relavent rotating frame, $\hat{\bf c} = \sqrt{\gamma} ({\hat{\sigma}}_1, {\hat{\sigma}}_2, {\hat{\sigma}}_3)$, and 
$d{\bf V} = (dV_1, dV_2, dV_3)^{\top}$. An unraveling equivalent to the one described by the SME given in \erf{SME:M3} can 
be realized for a quantum system with one environment (for example, in a circuit QED setup) but by measuring the three Pauli
 observables concurrently \cite{RusKorMol10,RusComWisMol12}.


 We consider the case where \blk the complex Wiener processes are taken to be 
in the form of \erf{W:ML} with $L = 3$.  Specifically, we consider performing 
three different unravelings, labelled by $k \in \{1,2,3\}$, in which the experimenter obtains
 information about $\hat\sigma_k$ from the \blk   by choosing $\phi_k=0$ for the $k$th component of the 
observable, while giving rise to noisy Hamiltonian evolution  for the other components by 
choosing $\phi_l=\pi/2$ for the $l\neq k$. \blk Every one of
 these three monitorings can be encapsulated in an unraveling matrix as shown in \erf{formU}. For instance,
 for the $k=3$ unraveling it is simplified to $\Theta = \eta I,$ and $ \Upsilon = \eta \, {\rm diag}(-1, -1, 1)$.
   The intuition behind this choice is that each unraveling yields information only about one 
  observable, and so would be expected to make this observable take a well-defined value. This 
would lead to quite different conditioned states for the different unravelings, which is what 
is needed to disprove all OPDMs.\blk 

\begin{figure}
\includegraphics[scale=0.60]{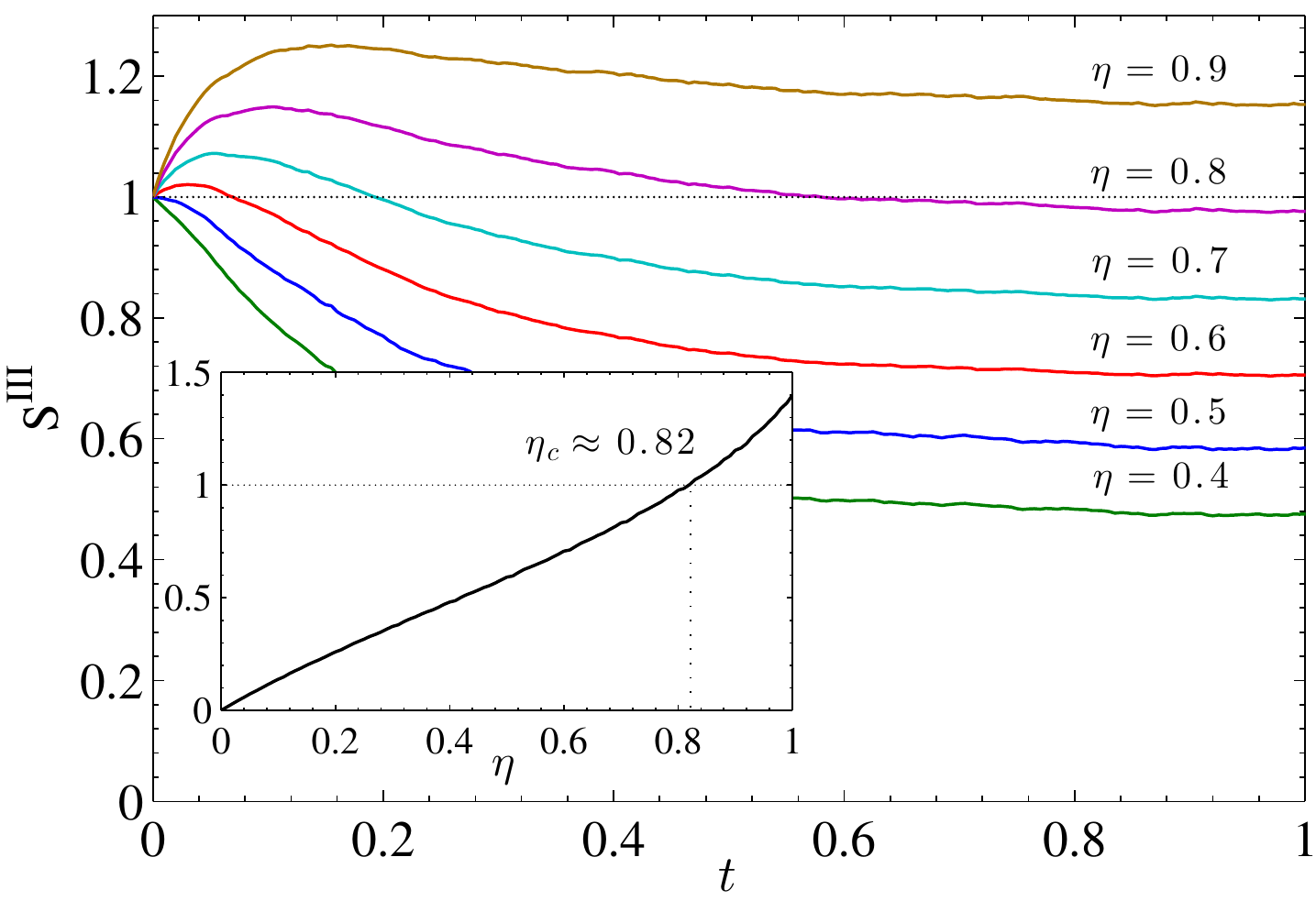}
\caption{\label{fig:M3} (Color online). Steering parameter for $K=3$ measurement settings: the temporal behavior of the steering 
parameter $S^{\rm III}$, \erf{S:M3},
for different values of the detection efficiency $\eta$. It is evident that all unravelings for which $\eta \leq 0.5$ are not able to demonstrate
EPR-steerability. Nevertheless, once the system settles in the steady state ($t \rightarrow \infty$), it still requires great improvements on detection
efficiency to the extent that it should roughly satisfy $\eta > 0.8$. Inset shows the value of $S^{\rm III}$ versus $\eta$ for the system in $\varrho_{\rm  ss}$.}
\end{figure}


 Thus \blk to prove the detector-dependence of $\varrho$ a suitable EPR-steering criterion should be examined. 
Such a criterion is often expressed in the form of an inequality. For example, from 
 the fact that \blk the Bloch 
vector ${\bf r}=\an{(\s{1},\s{2},\s{3})\tp}$ 
 must obey $\| {\bf r} \|^2 \leq 1$, and the 
convexity of $\an {{\hat{\sigma}}_l} \equiv {\rm Tr}[\varrho {\hat{\sigma}}_l]$ in the conditional state matrix $\varrho$, 
 it can be shown that every OPDM satisfies \blk \cite{WisGam12,CavRei09}
\beq \label{S:M3}
S^{\rm III} \equiv \sum_{{ k}=1}^3 {\rm E}^{ k} \big[ {\an {{\hat{\sigma}}_{ k}}}^2 \big] \leq 1.
\eeq
Here ${\rm E}^k[\bullet]$ means the ensemble average under the unraveling $k$. That is, three unravelings are implemented to calculate
the steering parameter $S^{\rm III}$. Note that due to the symmetrical nature of the problem the ensemble average
does not depend on ${ k}$ when the system is in the steady state for $t \rightarrow \infty$. 

Details of calculations are presented in \arf{appnA1}. As a function of time $t$ we plot in \frf{fig:M3} the steering
 parameter given in \erf{S:M3} for various values of the detection efficiency 
$\eta$. It can be clearly seen that as long as $\eta \leq 0.5$ the EPR-steering inequality is never violated. This should not be
very surprising as it was comprehensively analyzed in the previous section. Nevetheless, violation occurs while the system is in the 
transient state provided that $\eta > 0.5$. In contrast, once the system relaxes to its steady state (which might be much more interesting from practical 
points of view) only if $\eta > 0.8$ it is possible to establish EPR-steerable states. This is also illustrated in the inset where the value of 
$S^{\rm III}$ versus $\eta$ is drawn when the system dynamics reach to a stable point. The critical efficiency of $\eta_c \approx 0.82$ is required
to prove detector dependency of $\varrho$  in the steady state. \blk 

\begin{figure}
\includegraphics[scale=0.59]{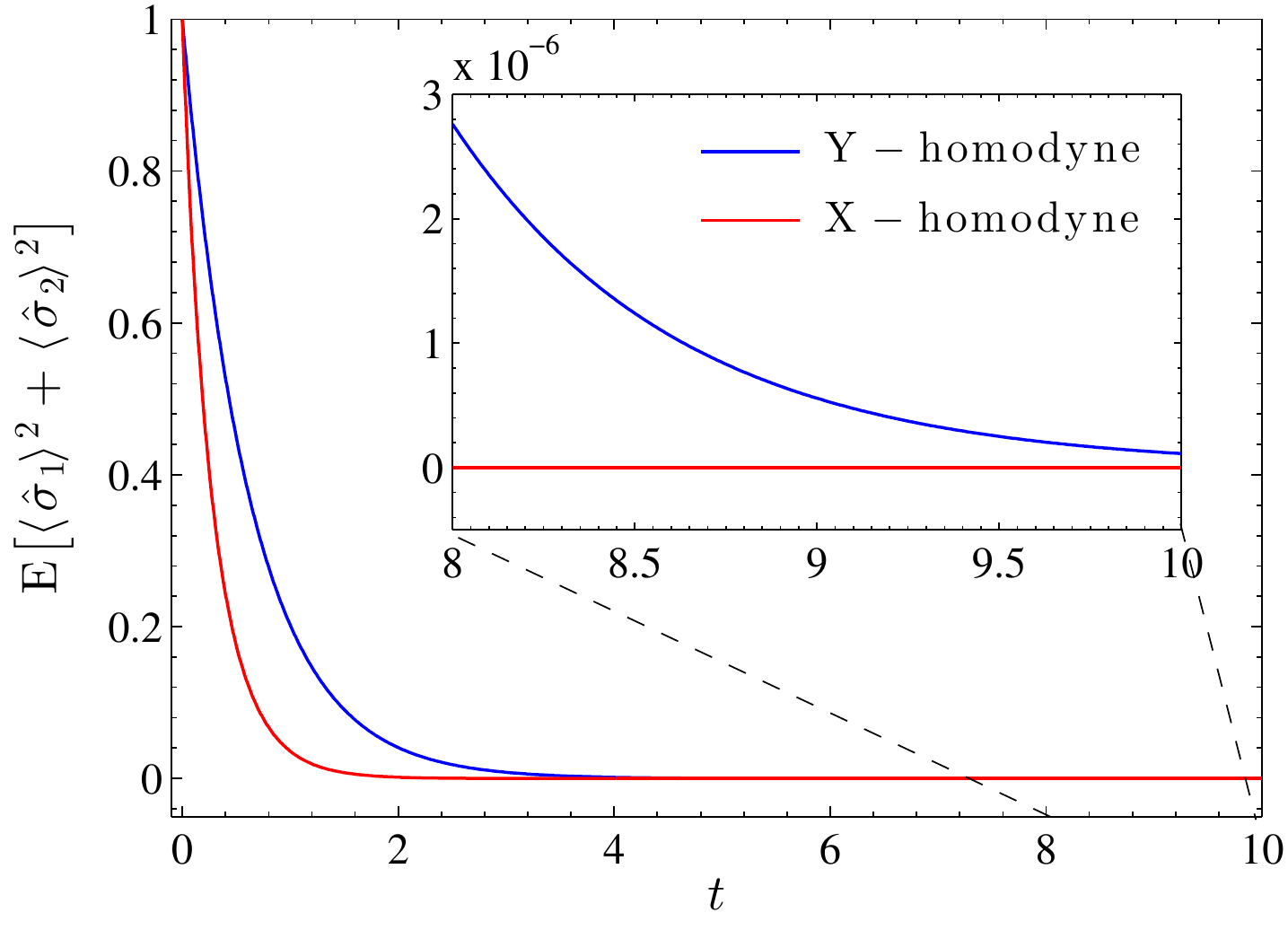}
\caption{\label{fig:M2Opt} (Color online). Ensemble average of 
$\big[ {\an {{\hat{\sigma}}_1}}^2 + {\an {{\hat{\sigma}}_2}}^2 \big]$: results of 
numerical simulations of the first term in the L.H.S of the inequality given in \erf{S:M2} using 
X- and Y-homodyne detection as a function of time for $\eta = 0.6$.}
\end{figure}

\subsection{$L=1$, $K=2$}  \label{sec:IV2}
The high efficiency required by the 3-measurement setting suggests  looking at the other extreme in which an open quantum system
decoheres just to a single environment. Consider that the SME of such a system is given by 
\beq \label{SME:M2}
{d\varrho} = {\cal D}[\sqrt{\gamma}{\hat{\sigma}}_3] \varrho\, dt+ {\cal H} [\sqrt{\gamma}\,{\hat{\sigma}}_3 dV] \varrho, 
\eeq
where again in notation of \erf{SME} Hamiltonian becomes zero in a suitable rotating frame, $\hat{\bf c} = \sqrt{\gamma} {\hat{\sigma}}_3$, and 
the Wiener process is given by \erf{W:ML} with $L=1$.  Here we consider just two different measurement strategies, 
corresponding to $\phi = 0$ (X-homodyne, giving maximum information), 
and $\phi = \pi/2$ (Y-homodyne, giving pure noise).  Since $L=1$, 
the correlation matrices are just scalars, $\Theta = \eta = \pm \Upsilon$ for the two respective cases. 

For the 2-measurement settings we study a class of EPR-steering inequalities similar to the one given
 in \erf{S:M3} which in general can be formulated as
\beq \label{S:M2Gen}
 {\rm E}^{\rm X} \Bigg[ \sum_{k=1}^3 \alpha_k {\an {{\hat{\sigma}}_k}}^2 \Bigg] + {\rm E}^{\rm Y} \Bigg[ \sum_{k=1}^3 (1 - \alpha_k) {\an {{\hat{\sigma}}_k}}^2 \Bigg] \leq 1,
\eeq
where 
\beq \label{alphak}
\forall k, \qquad  0 \leq \alpha_k \leq 1.
\eeq
This shows that there are an infinite number of such inequalities. However, one should carefully 
choose an optimal one so that the left hand side of \erf{S:M2Gen} constitutes a maximum. This is determined by 
whether the X-homodyne or Y-homodyne measurement scheme gives a larger value for the ensemble average of ${\an {{\hat{\sigma}}_k}}^2$.
If the former is the case one should choose 
$\alpha_k = 1$, and in the case of the 
latter $\alpha_k = 0$. There are protocols in which both X- and Y-homodyne detection give an equal amount of
information about the $k$th observable. In such a case $\alpha_k \in [0,1]$ is arbitrary. Note that the comparison between
${\rm E}^{\rm X} [{\an {{\hat{\sigma}}_k}}^2]$ and ${\rm E}^{\rm Y} [{\an {{\hat{\sigma}}_k}}^2]$ can be done through either 
numerical simulation or analytical calculation where possible.

In this work we have chosen the optimal  EPR-steering inequalities of the form (\ref{S:M2Gen})
by analyzing the numerical 
results. It turns out that for the SME given in \erf{SME:M2} the following inequality is optimal 
\beq \label{S:M2}
S^{\rm II} \equiv {\rm E}^{\rm Y} \big[ {\an {{\hat{\sigma}}_1}}^2 + {\an {{\hat{\sigma}}_2}}^2 \big] + {\rm E}^{\rm X} \big[ {\an {{\hat{\sigma}}_3}}^2 \big] \leq 1.
\eeq
This can be clearly seen in \frf{fig:M2Opt} where the ensemble average of ${\an {{\hat{\sigma}}_1}}^2 + {\an {{\hat{\sigma}}_2}}^2$ 
is plotted using both X- and Y-homodyne monitorings. The same argument holds for the second term of \erf{S:M2} but this time 
unraveling using X-homodyne gives more information about ${\hat{\sigma}}_3$ 
in comparison to Y-homodyne.
Each one of these two terms in \erf{S:M2} can be
realized in an an EPR-steering experiment as follows. Bob randomly chooses between the two measurement settings, 
and so tells Alice to implement either $\rm X$- or $\rm Y$- homodyne. 
 Bob then \blk correlates the outcomes of  his tomographic \blk measurements 
with those obtained  by Alice. For details, see Ref.~\cite{WisGam12}. \blk

\begin{figure}
\includegraphics[scale=0.6]{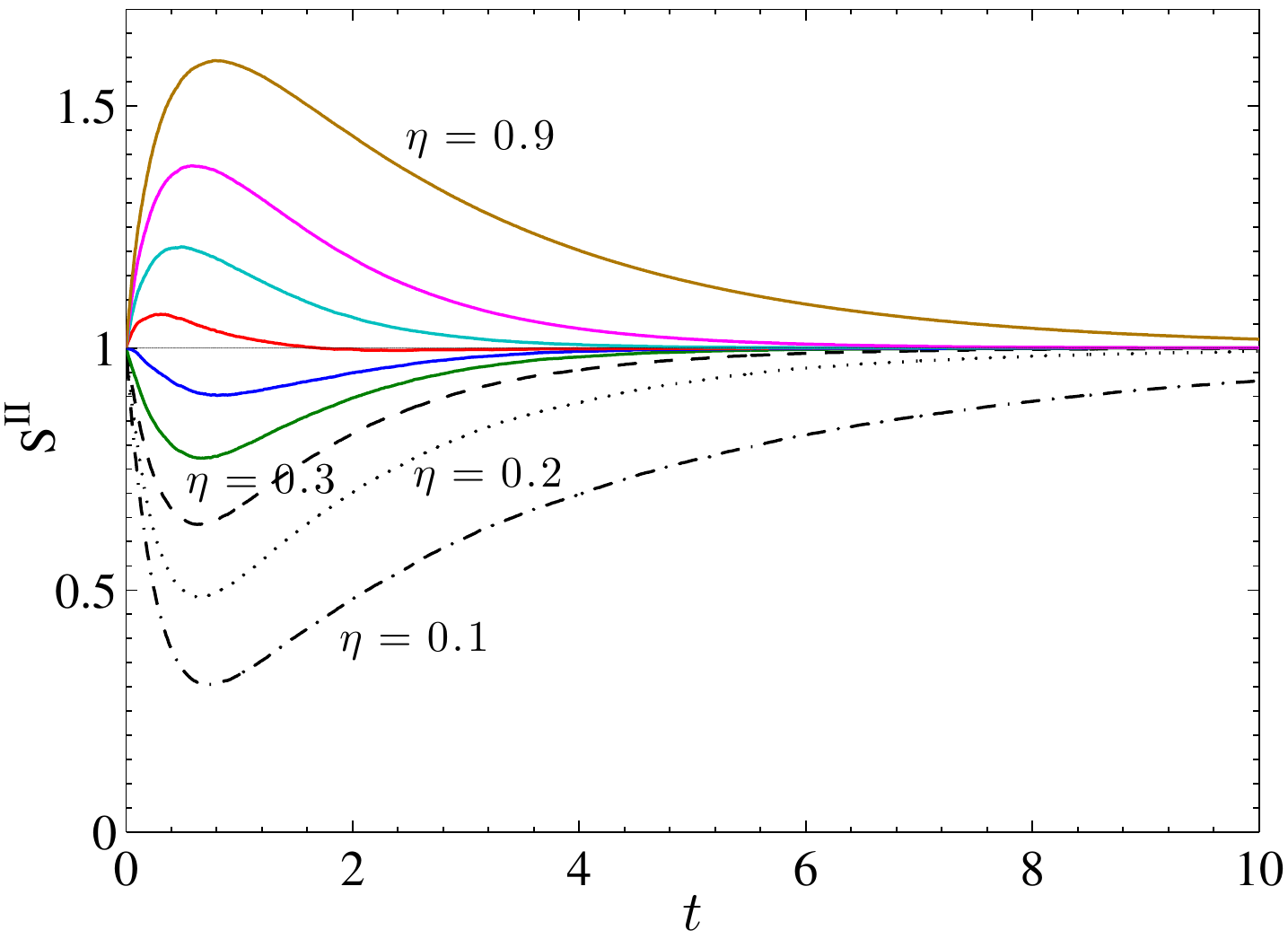}
\caption{\label{fig:M2} (Color online). Steering parameter for $K=2$ measurement settings: results of numerical simulations 
of the steering parameter $S^{\rm II}$, \erf{S:M2},
as a function of time for different values of $\eta$. Here for the sake of clarity we discard labels in 
between solid curves, but they follow the same color style 
and order as in \frf{fig:M3}. When the detection efficiencies satisfy $\eta \leq 0.5$ there is no possibilities (for 
 homodyne unravelings) to demonstrate EPR-steering, 
 as expected. However, beyond this limit,  it is possible to observe EPR-steering 
if the system is in its transient phase of evolution. In contradiction to the  case of $L=3$, $K=3$, in the long-time limit there is 
never any violation of \blk \erf{S:M2}. Curves for $\eta \le 0.3$ illustrate that the time taken to reach the asymptotic value of 1 
grows as the efficiency gets smaller.}
\end{figure}

In \frf{fig:M2} the time variation of the steering parameter $S^{\rm II}$ for different values
of detection efficiency $\eta$ is shown. There are two key features closely resembling those of 
 Sec.~\ref{sec:IV1}. \blk Firstly,
the necessary condition $\eta > 0.5$ for ruling out all OPDMS is  satisfied. \blk Secondly, the transient phase of evolution
establishes EPR-steering states provided the detection efficiency is greater than 0.5. 

 However, in a third respect 
this case is quite different: in the long-time limit it is impossible to prove detector dependency of $\varrho$
no matter how close to unity $\eta$ is. \blk 
This is because the $\rm Y$- homodyne unraveling generates 
just pure noise,  which contains information about the system evolution (the noisy Hamiltonian) but 
no information about the system state. Thus for any $\eta < 1$, the system purity decays deterministically 
over time under the $\phi=\pi/2$ unraveling, eventually reaching a completely mixed state as $t\to \infty$. In this limit, 
the first term in \erf{S:M2} will be zero, and since the second term is bounded above by unity, there is no 
way the inequality can be violated. Violations occur at short times only because we started the system 
in a pure state. \blk 
Further details are given in \arf{appnA2}.  It is interesting to \blk
note that the transient evolution here spans over an interval  several \blk times longer
than the duration in which   {$S^{\rm III}$ attains \blk steady state. 


\subsection{$L=2$, $K=2$}  \label{sec:IV3}

 Finally, we concisely present the results of our studies 
 of an open qubit system
coupled to two irreversible channels ($L=2$), and  compare them to the two results 
of us previously obtained in ~\crf{DarWis14}. To this end consider a qubit SME in the form of
\beq  \label{SME:M2L2}
d{\varrho} = \sum_{l=\pm} \Big({\cal D}[\sqrt{\gamma_l}{\hat{\sigma}}_l] \varrho + {\cal H} [\sqrt{\gamma_l}{\hat{\sigma}}_l dV_l] \varrho \Big),
\eeq
where $\s{\pm} = (\s{1}\pm i \s{2})/2$  are raising and lowering operators. 
As before,  the system's conditional state matrix $\varrho$ evolves in an appropriate rotating frame so that $\hat{H}=0$ in the notation of
\erf{SME}. Also the  Lindblad vector is $\hat{{\bf c}} = (\sqrt{\gamma_-} \s{-}, \sqrt{\gamma_+} \s{+})$, and
the independent Wiener processes are given by \erf{W:ML}. Thus, depending on the measurement strategy, determined 
by the phase term $\phi$ in Wiener process, the average current of the two different decoherence channels would be
\beq  \label{J:M2L2Gen}
\an{J_{\pm}} = \eta \sqrt{\gamma_{\pm}} \an{{\hat{\sigma}}_{\pm} + e^{i 2 \phi} \, {\hat{\sigma}}_{\mp}}, 
\eeq
such that the X-homodyne detection ($\phi=0$) gives information about ${\hat{\sigma}}_1$
\beq  \label{J:M2L2X}
\an{J_{\pm}}^X = \eta \sqrt{\gamma_{\pm}} \an{{\hat{\sigma}}_1}, 
\eeq
and the outcome of the Y-homodyne detection ($\phi=\pi/2$) is given by
\beq  \label{J:M2L2Y}
\an{J_{\pm}}^Y = {\pm} i  \eta \sqrt{\gamma_{\pm}} \an{{\hat{\sigma}}_2}.
\eeq  

This SME has been studied in \crf{DarWis14} in order to propose an experimental test for disproving all OPDMs using quantum 
jumps unravelings for experimental setups with detection efficiencies less than $50\%$. In this protocol, however, Alice
should be able to realize $n$ measurement settings in the azimuthal direction $\varphi \in \{ (j/n) \pi \}_{j=1}^{n}$ and 
one in the $z$ direction to be able to work out the considered EPR-steering inequality \cite{DarWis14} 
\beq  \label{S:MnL2}
  \frac{1}{n}\sum_{j=1}^n {\rm E}^{\varphi_j} \Big[\abs{\an{\hat{\sigma}_{\varphi_j}}}\Big]
  -  f(n){\rm E}^z\sq{ \sqrt{1-{\langle {\hat{\sigma}}_{3} \rangle}^2}} \le 0 .
\eeq 
Here $\s{\varphi}=\s{-}e^{i\varphi}+\s{+}e^{-i\varphi}$, and the function $f(n)$ is given in Ref.~\cite{JonWis11} which monotonically decreases with $n$, asymptoting to $2/\pi$ as $n\to \infty$. Thus in the limit of large $n$  
a smaller detection efficiency is required to demonstrate detector dependence of the conditional state of an open quantum system. It turns out that 
the critical efficiency is approximately $\eta_c \approx 0.59$ provided that one is capable of performing 
an infinite number of monitoring schemes, and in addition the condition $R \equiv \gamma_+/\gamma_- \ll 1$ is 
satisfied. These assumptions are nontheless not very convenient. The former is not practical as it requires great 
number of measurement settings and the latter leads to a very small violation in comparison with the maximum 
possible violation when the detection is perfect and  $R = 1$ \cite{DarWis14}. 
However, it was shown that with just a finite number of measurement settings, $K = n + 1 = 5$, one would be able to have a decent 
(0.05) violation of \erf{S:MnL2} with $\eta \approx 0.78$. 
In what follows we show that an even more suitable EPR-steering 
inequality can be derived to rule out all OPDMs with a sufficient condition as close as possible (from above) to 
the necessary condition bound of 50\%.   

Thus for the case given by \erf{SME:M2L2}, because the $\hat\sigma_\pm$ are not Hermitian, 
both the X- and Y-homodyne measurements yield information about the system, but about different observables,  
namely $\hat \sigma_1$ and $\hat \sigma_2$ respectively, given in \erfa{J:M2L2X}{J:M2L2Y}. This suggests yet 
another variation derivable from the 
inequality $\| {\bf r} \|^2 \leq 1$, namely the following EPR-steering inequality 
\beq \label{S:M2L2}
S_{\pm}^{\rm II} \equiv {\rm E}^{\rm X} \big[ {\an {{\hat{\sigma}}_1}}^2 + \frac{{\an {{\hat{\sigma}}_3}}^2}{2} \big] + {\rm E}^{\rm Y} \big[ {\an {{\hat{\sigma}}_2}}^2 + \frac{{\an {{\hat{\sigma}}_3}}^2}{2} \big] \leq 1,
\eeq
which is a special case of \erf{S:M2Gen} with $\alpha_1 = 1, \alpha_2 = 0$ and $\alpha_3 = 1/2$. This 
assignment should be now obvious considering the argument made in \srf{sec:IV2} and recalling \erfa{J:M2L2X}{J:M2L2Y}. 
In fact the value of $\alpha_3$ is arbitrary as both 
X- and Y-homodyne give the same amount of information about $\hat{\sigma}_3$. The choice  
$\alpha_3 = 1/2$ emphasizes the symmetry. \blk 

For sufficiently small $R \ll 1$ it is easy
to derive an analytical relation for the steering parameter in \erf{S:M2L2},
\beq  \label{S:analytic}
S_{\pm}^{\rm II} \cong 1 + 8 (\eta - 0.5) R.  
\eeq
The details of the calculation are presented in \arf{appnA3}.
The analytic relation, \erf{S:analytic}, cries out 
not only the necessary condition mentioned in the \srf{sec:III}, 
but also shows that $\eta > 0.5$ is sufficient to rule out all OPDMs. However, this result holds only for small $R$'s and as such 
it will not be convincible enough from experimental viewpoint due to a very small violation. Therefore, a desired objective would 
be to achieve as low as possible to $\eta_c \approx 0.5$ and obtain a decent 
violation of \erf{S:M2L2} so that realizing experimentally EPR-steerable states, and hence 
To this end, for a decent violation of \erf{S:M2L2} an optimal value of $R$ is obtained. For example, a few percent of 
the difference between the threshold and the maximum possible value allowed by quantum mechanics (here unity and 2, respectively) 
should be almost enough to take into account imperfections. It turns out that an efficiency of approximately $0.72$ is 
needed to acquire a 5\% violation with the ratio of transition rates being optimized at $R \approx 0.2$, see \frf{fig:M2L2}. In 
this figure we plot $S^{\rm II}_{\pm}$ as a function of $\eta$ with the critical efficiency $\eta_c = 0.68$ for violating \erf{S:M2L2}. 
Also, the inset (bottom-right) depicts the same steering parameter versus $R$ for a fixed detection efficiency $\eta = 0.72$. 
\begin{figure}
\includegraphics[scale=0.59]{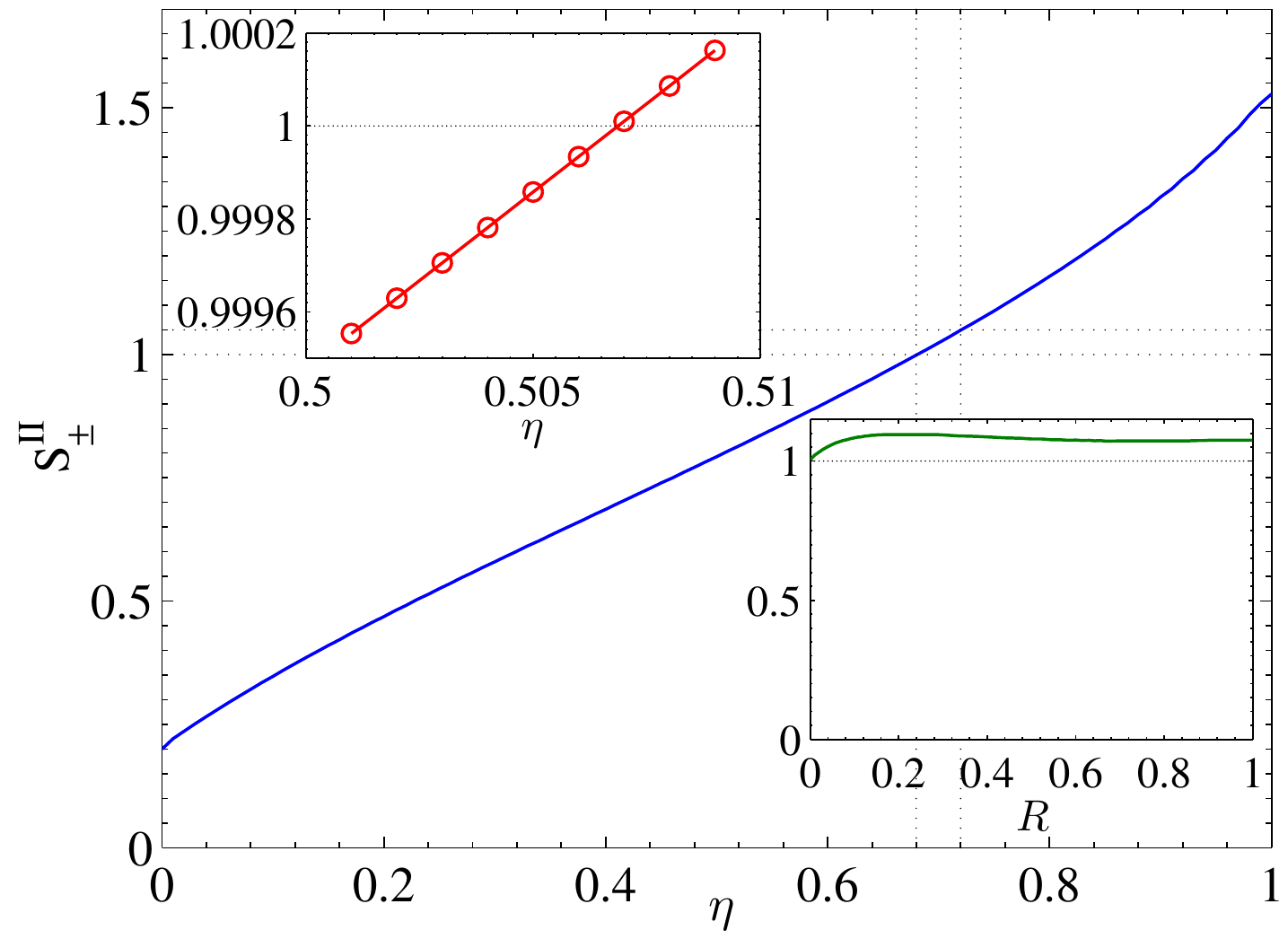}
\caption{\label{fig:M2L2} (Color online). Steering parameter for $K=2$ measurement settings and $L=2$ decoherence 
channels into the environment: we plot $S^{\rm II}_{\pm}$, \erf{S:M2L2}, as a function of $\eta$ for an optimal value of $R\approx 0.2$ with
a significant violation of 5\% at $\eta = 0.72$; see text for details. Critical efficiency here is $\eta_c = 0.68$. Inset: top-left shows the same plot but for $R \approx 0.01$ with critical efficiency of $\eta_c \approx 0.507$; bottom-right displays variation of $S^{\rm II}_{\pm}$ versus $R$ for $\eta=0.72$.}
\end{figure}

Finally, one might question the validity of approximations made to obtain \erf{S:analytic}, in the sense that it is 
implying that $\eta > 0.5$ should be sufficient for disproving all OPDMs in the limit where $R$ is very small. To support
 these simplifications which led us achieve the above mentioned analytic formula we present the outcome of
numerical simulations of \erf{S:M2L2} for $R \approx 0.01$ in the inset (top-left) of \frf{fig:M2L2}. As it can be seen $S_{\pm}^{\rm II}$
varies linearly with respect to $\eta$ and this is exactly the behavior one would expect from \erf{S:analytic}. Moreover, this plot 
demonstrates that the critical efficiency is $\eta_c \approx 50.7\%$ which is very close to what analytical calculation predicts,
that is $50\%$. This is a very good improvement on the critical efficiency required for proving detector dependence of stochastic
conditional states $\varrho$. 

\section{Conclusion}     \label{sec:V}
In summary, we have investigated the dependence of diffusive unravelings of open quantum systems on the existence of a 
distant detector. Due to imperfect detection schemes some different measurement strategies on an open qubit system were 
studied to obtain a minimum quantum efficiency required to disprove objectiveness of pure-state evolutionary models. 
In this work we sought two objectives. Firstly, to determine some general 
necessary conditions for proving detector dependency of dynamical quantum events.And secondly, to derive 
sufficient conditions by considering different protocols.

We derive necessary conditions for being able to demonstrate EPR-steering 
in the form of a no-go for inefficient diffusive unravelings. These conditions depend only on 
the parameters of the unravelings, not on any details of the system. This 
theorem can be applied in its most general form by recasting the condition 
as a special type of semidefinite programming known as a 
feasibility problem. The simple case of $L=1$ can be thoroughly analyzed by a graphical representation 
involving right cones. 

In contrast to the case of necessary conditions, deriving sufficient conditions can only be 
done case by case, in that they depend on the systems being examined. To achieve this
goal three factors may be inspected for an open qubit system: varieties of irreversible evolution equations, 
different measurement schemes, and various
EPR-steering inequalities. The threshold efficiencies required $\eta_c$ for each scenario along with those of 
other previously proposed experimental tests
(for diffusive unravelings) are summarized in \trf{tbl:S}. The investigated environmental decoherence channels fall into 
two categories; dissipateve (Diss) and non-dissipative (ND). According to the results obtained in this work, the former 
class shows better figures for the critical efficiency. Among all of these the case in which the system 
decoheres into two environments 
and the latter undergo monitoring via two measurement 
settings (\srf{sec:IV3})} shows a promising candidate for being probed 
empirically. Lastly, this work may, on the one hand, encourage experimentalists to 
endeavour to increase the efficiencies of detection schemes, and, on the other hand, stimulate theorists 
to invent still more robust EPR-steering tests. 


\begin{table}
\caption{\label{tbl:S} The critical efficiencies $\eta_c$ required for violating different EPR-steering criteria. Except 
for the last two rows, the rest represent entirely different physical systems coupled to different number of 
decoherence channels; dissipative (Diss) and non-dissipative (ND). Also for each case specific 
measurement schemes are considered.}
  \begin{tabular}{ c  c  c  c  }
  \hline \hline
    \begin{tabular}{@{}c@{}}Decoherence \\ channels ($L$ )\end{tabular} &  \begin{tabular}{@{}c@{}}Measurement  \\ settings ($K$ )\end{tabular} & \begin{tabular}{@{}c@{}}Steering \\parameter\end{tabular} &  \begin{tabular}{@{}c@{}} $ t\to \infty$ \\ $\eta_c$ \end{tabular}  \\ \hline
    3 (ND) & 3 & $S^{\rm III}$, \erf{S:M3} & \;\; 0.82   \\ 
    1 (ND) & 2 & $S^{\rm II}$, \erf{S:M2} & \;\; --  \\ 
    1 (Diss) & 2 & Eq.(3) of \crf{WisGam12} & \;\; 0.73 \\
    2 (Diss) & 2 & $S^{\rm II}_{\pm}$, \erf{S:M2L2} & \;\; 0.5 \\
    2 (Diss) & 2 & Eq.(8) of \crf{DarWis14} & \;\; 0.59 \\
     \hline \hline
  \end{tabular}
\end{table}


\section*{Acknowledgement}

\input{acknowledgement.tex }

\appendix
\numberwithin{equation}{section}
\section{Simulating averages} \label{appnA} 
\subsection{$L=3$, $K=3$}  \label{appnA1} 
For the three measurement settings general dyne unraveling as mentioned in the main text, depending on the 
measurement strategy, \erf{formU} is written in the following forms
\bqa 
U_{|_{k = 1}} & = & \eta\, {\rm diag}(1,0,0,0,1,1), \\
U_{|_{k = 2}} & = & \eta\, {\rm diag}(0,1,0,1,0,1), \\
U_{|_{k = 3}} & = & \eta\, {\rm diag}(0,0,1,1,1,0), 
\eqa
which are clearly PSD. Then the conditional state of the system which
evolves according to \erf{SME:M3} can be rewritten in the Bloch representation so that dynamics of the system is governed by the 
following multi-dimensional stochastic differential equation (MDSDE)
\beq \label{sde:M3}
d{\bf r} = {\bf A}({\bf r}) dt + {\bf B}({\bf r}) d{\bf W}(t), 
\eeq
where 
\bqa 
{\bf r} & = & (x, y, z)^\top \equiv (\an {{\hat{\sigma}}_1}, \an {{\hat{\sigma}}_2}, \an {{\hat{\sigma}}_3})^\top, \label{r:M3} \\
{\bf W} & = & (W_1, W_2, W_3)^\top, \label{W:M3}\\
{\bf A} & = & - 4 {\bf r}, \label{A:M3}
\eqa
and the $W_j$s are independent Wiener processes. The matrix ${\bf B}$ also takes the following form
\beq  \label{B:M3}
{\bf B} = 2 \sqrt{\eta} \left[ 
\begin{array}{ccc}
0  & -z & - x z \\ 
z  &  0 & - y z \\ 
-y  & x & 1 - z^2 
\end{array}
\right].
\eeq
Here, we have just expressed details of \erf{sde:M3} for the $k = 3$ case. Note that the ultimate goal is to calculate 
the steering parameter $S^{\rm III}$, given in \erf{S:M3}, composed of ensemble averages. By virtue of
symmetrical nature of the problem the ensemble average ${\rm E}^k [ {\an {{\hat{\sigma}}_k}}^2 ]$ is independent 
of $k$ when the system evolves, for a sufficiently long time. Also note that the $k$th measurement strategy merely gives 
noisy output for the rest of observables. That is, ${\rm E}^k [ {\an {{\hat{\sigma}}_{k^\prime}}}^2 ] = 0$ 
for $k^{\prime} \neq k$.

As it was discussed in \srf{sec:IV1} the $k$th unravelling of the SME, \erf{SME:M3}, gives only 
information about the $\hat\sigma_k$ observable. Therefore, to calculate ${\rm E}^k \big[ {\an {{\hat{\sigma}}_k}}^2 \big]$ in \erf{S:M3} it is 
convenient to introduce some variables $\beta, \theta$ and $\gamma$ with the following relations
\bqa  \label{polcor}
{\an {{\hat{\sigma}}_m}} + i {\an {{\hat{\sigma}}_n}} & = & \sqrt{\beta} e^{i\theta}, \quad  {\rm for} \quad m, n \neq k \label{polcorB} \\
{\an {{\hat{\sigma}}_k}}^2 & = & \gamma.  \label{polcorG}
\eqa
This allows us to work with a MDSDE which is more relevant to what we need to work out \erf{S:M3}. The following lists different matrices
that characterize such a MDSDE for $k = 3$
\bqa 
{\bf {\ut r}} & = & (\beta, \theta, \gamma)^\top, \label{r:M3pol} \\
{\bf {\ut W}} & = & (W_\beta, W_\theta, W_\gamma)^\top, \label{W:M3pol}
\eqa
and
\bqa  
{\bf A} & = & \left[ 
\begin{array}{c} \label{A:M3pol}
8 \eta \gamma + 4 \eta \beta \gamma - 8 \beta   \\[2pt] 
\frac{1}{2} \eta \gamma^2 \sin(4 \theta)  \\ [2pt]
4 \eta \beta + 4 \eta (1 - \gamma)^2 - 8 \gamma
\end{array}
\right], \\[4pt]
{\bf B} & = & 4 \sqrt{\eta} \left[ 
\begin{array}{ccc} \label{B:M3pol}
-\sqrt{\beta \gamma}  & 0 & - \beta \sqrt{\gamma} \\ 
0  &  \sqrt{{\gamma}/{(4\beta)}} & 0 \\ 
\sqrt{\beta \gamma}  & 0 & \sqrt{\gamma} (1 - \gamma) 
\end{array}
\right].
\eqa

Therefore, to evaluate $S^{\rm III}$ in \erf{S:M3} the ensemble average of $\gamma$ for each individual unraveling, 
${\rm E}^k \big[ {\an {{\hat{\sigma}}_k}}^2 \big]$, should be calculated. As it follows from \erfs{r:M3pol}{B:M3pol}, the evolution 
of $\gamma$ in turn is coupled to that of $\beta$ and is independent of variations of $\theta$. Thus, a set of just two coupled SDEs
for $\beta$ and $\gamma$ is simulated. This MDSDE is solved numerically using the Milstein method \cite{KloPla92}. For 
sufficiently long time that the system is in the steady state data are sampled and the ensemble averages are calculated. 

\subsection{$L=1$, $K=2$} \label{appnA2}
The unraveling matrix based upon homodyne measurements on the output field of the system described by \erf{SME:M2} 
becomes $U = \eta \, {\rm diag}(1,0)$. Then the stochastic equations of motion in the notation of \erfs{sde:M3}{W:M3} can be expressed as 
\beq  \label{AB:M2}
{\bf A} = -2 \left[ 
\begin{array}{c}
 1\\ 
 1\\ 
0 
\end{array}
\right], \quad
{\bf B} = 2 \sqrt{\eta} \left[ 
\begin{array}{ccc}
0  & 0 & - x z \\ 
0  & 0 & - y z \\ 
0  & 0 & 1 - z^2 
\end{array}
\right].
\eeq
Now using \erfs{polcor}{W:M3pol}, this MDSDE is transformed into 
\bqa  \label{AB:M2pol}
{\bf A} & = & 4 \eta \left[ 
\begin{array}{c} \label{A:M2pol}
\gamma \beta - \beta/\eta\\[4pt] 
\frac{\gamma}{8} {\rm sin}(4\theta)\\ [4pt]
(1-\gamma)^2
\end{array}
\right], \\
{\bf B} & = & 4 \sqrt{\eta \gamma}  \left[ 
\begin{array}{ccc} \label{B:M2pol}
0  & 0 & - \beta \\  
0  & {1}/{\sqrt{4\beta}} & 0 \\ 
0  & 0 & 1 - \gamma 
\end{array}
\right].
\eqa
The same argument as the three measurement settings holds here where the system dynamics at any given time can be
described by variables $\beta(t)$ and $\gamma(t)$, and also the same procedure is employed to calculate the ensemble average of 
$\big[ {\an {{\hat{\sigma}}_{3}}}^2 \big]$. 

For the noisy Hamiltonian evolution measurement scheme though, the unraveling matrix is 
$U = \eta \, {\rm diag}(0,1)$. The ensemble average of $\big[ {\an {{\hat{\sigma}}_1}}^2+{\an {{\hat{\sigma}}_2}}^2 \big]$, which is
equal to that of $[\beta]$ through \erf{polcorB}, is obtained by solving the following SDEs
\bqa 
d\beta & = & 4 \beta \, (\eta - 1) \, dt, \\
d\gamma & = & 0,
\eqa
which lead to the equations of motion 
\bqa 
\beta(t) & = & \beta^{\rm }(0) \, e^{4 (\eta-1) t}, \label{b:M2NK} \\
\gamma(t) & = & \gamma^{\rm }(0). \label{g:M2NK}
\eqa

\subsection{$L=2$, $K=2$} \label{appnA3}
In this scenario the homodyne detections (with phases $\phi$ and $- \phi$) of the two environmental channels gives the following 
correlation matrices
\beq
\Theta = {\rm diag} (\eta, \eta), \quad \Upsilon = \eta \,{\rm diag}(e^{-2i\phi}, e^{2i\phi}).
\eeq
These diffusive unravelings yield information only about observables $\hat{{\sigma}}_1$ ($\phi = 0$) and 
$\hat{{\sigma}}_2$ ($\phi = \pi/2$). In fact, one of 
the reasons why we are working with the specific EPR-steering inequality given in \erf{S:M2L2} is because there is not an 
appropriate unraveling to be useful for calculating the ensemble average of ${\an{\hat{{\sigma}}_3}}^2$. However, as it is the case here, 
one could still estimate it from the measurement record for $\hat{{\sigma}}_1$ and $\hat{{\sigma}}_2$.

Due to the symmetry of the problem both terms in the L.H.S of \erf{S:M2L2} return the same value. This means we 
can evaluate one and double it. Thus, the case of X-homodyne with $\phi = 0$ is studied. The dynamics of the qubit is 
confined to x-z plane such that at any instant of time its state can be described by a point $\big(x(t),0,z(t)\big)$ in the 
Bloch sphere, where the Bloch vector is ${\bf r} = (x, y, z)$. The trajectory of this point in the sphere is tracked by solving the coupled SDEs 
\bqa
 dx & = & -(\gamma_{\Sigma}/2) \, x dt + \sqrt{\eta\gamma_-} (1+z-x^2) dW_- + \nonumber \\[4pt] 
      & & \sqrt{\eta\gamma_+} (1-z-x^2) dW_+ ,  {\label {d1}}\\[4pt]
 dz & = & (-\gamma_{\Sigma} \, z + \gamma_{\Delta})  dt - \sqrt{\eta\gamma_-} x (1+z) dW_- + \nonumber \\[4pt]
      &  & \sqrt{\eta\gamma_+} x (1-z) dW_+, {\label {d2}} 
\eqa 
where $\gamma_{\Sigma} = \gamma_+ + \gamma_- $ and $\gamma_{\Delta} = \gamma_+ - \gamma_-$. As we have 
done for previous simulating jobs we use Milstein method to numerically compute these equations. Once the system 
reaches to its steady state, $t \rightarrow \infty$, data is being recored to calculate ${\rm E}^{\rm X} \big[ {\an {{\hat{\sigma}}_2}}^2 + 1/2{\an {{\hat{\sigma}}_3}}^2 \big]$, and hence ${\rm E}^{\rm Y} \big[ {\an {{\hat{\sigma}}_1}}^2 + 1/2 {\an {{\hat{\sigma}}_3}}^2 \big]$, for 
each measurement scheme as described above. 

Finally, when $R \equiv \gamma_+/\gamma_- \ll 1$, the conditional state of the system under diffusive unraveling is almost in proximity of
the ground state so that approximations $x = O\big(\sqrt{R}\big)$ and $1+z = O(R)$ hold. Then \erfa{d1}{d2} are transformed, to leading order, to
 \begin{eqnarray}  
 dx & = &  -(\gamma/2) \, x\, dt + 2 \sqrt{\eta \gamma R} \,dW_+ ,    {\label {approx1}}\\[4pt] 
dz & =  & \gamma [ 2R - z - 1]  dt  + 2 \sqrt{\eta \gamma R} \, x\, dW_+, {\label {approx2}}
 \end{eqnarray} 
 where $\gamma = \gamma_-$. 
It is straightforward to calculate the ensemble averages ${\rm E} [x^2] \approx 4 \eta R$ and ${\rm E} [z^2] \approx 1 - 4R$ for the steady state. Plugging these results back into \erf{S:M2L2} it is easy to show that for small $R$ it is approximated by
\beq
S_{R\ll1} \cong 1 + 8 (\eta - 0.5) R.
\eeq


\end{document}

%% file: author_list.tex
\author{Shakib Daryanoosh} \email{s.daryanoosh@griffith.edu.au}
\author{Howard M.~Wiseman} \email{h.wiseman@griffith.edu.au}
 \affiliation{Centre for Quantum Computation and Communication Technology (Australian Research Council),Centre for Quantum Dynamics, Griffith University, Brisbane, Queensland 4111, Australia}
 \author{Jay M.~Gambetta} \email{jay.gambetta@us.ibm.com}
 \affiliation{IBM T. J. Watson Research Center, Yorktwon Heights, New York 10598, USA}
 
%
 
%
%


%
\vskip 0.25cm

%% file: acknowledgement.tex
This research was supported by the ARC Centre of Excellence Grant No. CE110001027. 

%% file: Q_Diffusive_Monitoring.bbl
\begin{thebibliography}{99}
\bibitem{ft1}
Note that one could encode both matrices $\Upsilon_0$ and $\Theta_0$ in the vector ${\bf x}_0$. However, this not only expands the size of unknowns by $L$, which is undesirable, but also gives no additional information in comparison with the hypothetical case of $\Theta_0 = I$. That is, if there does not exit an unraveling $U(I, \Upsilon_0)$ by which all other unravelings can be coarse grained, then it would be entirely pointless to add another layer of complexity to the problem by considering $\Theta_0 \neq I$.
\end{thebibliography}
